\documentstyle[graphicx]{mn}

\newif\ifAMStwofonts
\AMStwofontstrue

\def\pg{{PG1211+143}}

\def\xmm{{\it XMM-Newton}}
\def\chandra{{\it Chandra}}
\def\suzaku{{\it Suzaku}}

\def\et{{et al.\ }}

\newcommand{\ls}{\mathrel{\hbox{\rlap{\hbox{\lower4pt\hbox{$\sim$}}}\hbox{$<$}}}}
\newcommand{\gs}{\mathrel{\hbox{\rlap{\hbox{\lower4pt\hbox{$\sim$}}}\hbox{$>$}}}}

\def\Msun{\hbox{$\rm ~M_{\odot}$}}

\def\rchi{{$\chi^{2}_{\nu}$}}

\def\H0{{\rm ~km~s^{-1}~Mpc^{-1}}}

\def\et{{et al.}}

\title[A soft X-ray wind]
        {Imprints of a high velocity wind on the soft x-ray spectrum of \pg}
\author[K.A.Pounds \et]
        {K.A.Pounds $^{1}$, A. Lobban $^{1}$, J.N.Reeves $^{2}$, S.Vaughan $^{1}$ and M. Costa $^{2}$ \\
$^{1}$ Department of Physics and Astronomy, University of Leicester, Leicester, LE1 7RH, UK \\
$^{2}$ School of Physical Sciences, Keele University, Keele, ST5 5BG, UK \\}

\date{Accepted for publication in MNRAS}
\pagerange{\pageref{firstpage}--\pageref{lastpage}}
\pubyear{2016}
\begin{document}
\maketitle
\label{firstpage}

\begin{abstract}
An extended \xmm\ observation of the luminous narrow line Seyfert galaxy \pg\ in 2014 has revealed a more complex high velocity wind, with components distinguished in velocity, ionization level, and
column density. Here we report soft x-ray emission and absorption features from the ionized outflow, finding counterparts of both high velocity components, v$\sim$0.129c and v$\sim$0.066c,
recently identified in the highly ionized Fe K absorption spectrum. The lower ionization of the co-moving soft x-ray absorbers imply a distribution of higher density clouds embedded in the main
outflow, while much higher column densities for the same flow component in the hard x-ray spectra suggest differing sight lines to the continuum x-ray source.
\end{abstract}


\section{Introduction}

The narrow-line Seyfert galaxy \pg\ is widely associated with ultra-fast, highly ionized outflows (UFOs), since the first detection in  a non-BAL (broad absorption line) AGN (active galactic nucleus) of
strongly blue-shifted absorption lines of highly ionized gas, corresponding to a sub-relativistic outflow velocity of 0.15$\pm$0.01c (Pounds \et\ 2003, Pounds and Page 2006).  Archival data from \xmm\
and \suzaku\ have since shown similar UFOs are relatively common in nearby, luminous AGN (Tombesi \et\ 2010, 2011; Gofford \et\ 2013). 

The frequency of these detections confirms a substantial covering factor - and hence significant mass and momentum in such winds,with the potential  to disrupt star formation in the host galaxy.
That realisation has led to the view (e.g. King 2003, 2010) that powerful AGN winds could explain the remarkable correlation of the supermassive black hole mass and the
velocity dispersion of the host galaxy's stellar bulge in a wide range of galaxies, the so-called M--$\sigma$ relation (Ferrarese and Merritt 2000, Gebhardt \et\ 2000).

It is noteworthy that the great majority of UFO detections (defined in Tombesi \et\ 2010 with an outflow velocity v $\geq$ 10000 km s$^{-1}$; 0.03c) are based on blue-shifted absorption lines being
identified with resonance transitions in  Fe XXV and XXVI ions. The high ionization state has made absorption features difficult to detect in the soft x-ray region covered by higher
resolution grating spectrometers on \xmm\ and \chandra. A recent exception came from a re-analysis of  the 2001 Reflection Grating Spectrometer (RGS: den Herder 2001) data of \pg, where a
wavelength-to-velocity transformation was used to combine Lyman-$\alpha$ absorption lines of several lower mass ions, finding compelling evidence for a soft x-ray UFO, with a Doppler-corrected velocity 
v$\sim$0.076c, a factor $\sim$2 less than that for the highly ionized flow (Pounds 2014). That lower velocity was confirmation of a second ionized
absorption component  required in a partial covering analysis of the combined 2001, 2004 and 2007 \xmm\ broad band spectra of \pg\ (Pounds and Reeves 2009), where modelling of continuum curvature made
the outflow velocity poorly constrained. 

To further explore the velocity and ionization structure of the fast wind in \pg, an extended \xmm\ observation was carried out during 7 spacecraft orbits between 2014 June 2 and 
July 9. On-target exposures for individual orbits ranged from $\sim$50 to $\sim$100 ks, with a total  duration of $\sim$630 ks. Full details of the \xmm\ observing log and data processing are given in
Lobban \et (2016), reporting the results of a detailed timing analysis. 

Pounds \et\ 2016 (hereafter Paper 1)  present an  analysis of the hard x-ray spectrum, based on data from the pn camera (Strueder 2001), revealing new spectral structure in Fe K
absorption lines, and dual velocities (v$\sim$0.066c and v$\sim$0.129c) in the highly ionized wind.  The present paper reports a corresponding analysis of the 2014 soft x-ray
spectrum of \pg\ using data from the RGS.

We assume an AGN redshift of $z=0.0809$ (Marziani \et\ 1996). Spectral fitting is based on the {\tt XSPEC} package (Arnaud 1996) and includes absorption  due to the line-of-sight Galactic column of
$N_{\rm H}$ =3$\times10^{20}\rm{cm}^{-2}$  (Murphy \et\ 1996, Willingale \et 2013).  90 \% confidence intervals on model parameters are based on  $\Delta\chi^{2}=2.706$. Estimates for the black hole
mass in \pg\ range from $3\times 10^{7}$\Msun\ (Kaspi \et\ 2000) to $1.5\times 10^{8}$\Msun\ (Bentz \et\ 2009), with the lower value making the historical mean luminosity close to Eddington. All
quoted velocities are corrected for  the relativistic Doppler effect.

\section{Soft X-ray absorption and emission from circumnuclear photoionized gas}

The composite 2014 RGS spectrum was produced by summing both RGS 1 and RGS 2 data over all 7 \xmm\ orbits with the SAS tool RGScombine, for a combined exposure of 1.27 Ms. For spectral modelling, the
data were binned to a minimum of 25 counts per bin compatible with use of the $\chi^{2}$ statistic, with the additional constraint that no bin be narrower than 1/3 of 
the mid-band FWHM resolution (Kaastra and Bleeker 2016). 

Before attempting to model soft x-ray absorption and emission spectra we first fit the underlying soft X-ray continuum by extending the continuum model found in Paper 1, consisting of a
hard and a soft power law, with the addition of a blackbody to match the soft excess. The soft power law component was required in Paper 1 by inter-orbit
difference spectra,  represented over the 2--10 keV spectral band by an unabsorbed power law of photon index $\Gamma$$\sim$2.9, and that continuum component was affirmed in the soft x-ray spectrum, finding
inter-orbit difference spectra described by a power law of index $\Gamma$$\sim$2.8$\pm$0.1. Such a steep continuum  component has the potential to substantialy dilute the imprint of photoionized gas on
the overall soft x-ray spectrum and is included here in modelling spectral features in the 
2014 RGS data.

Figure 1 illustrates the resulting continuum model for the stacked RGS data over the full 8--38 \AA\ waveband, with the corresponding data-model ratio indicating positive and 
negative residuals suggestive of significant absorption and
emission from circumnuclear ionized gas. A measure of that spectral structure is given by the fit statistic  \rchi\ of 1251/721.

\begin{figure}                                                                                                  
\centering 
\includegraphics[width=6.3cm, angle=270]{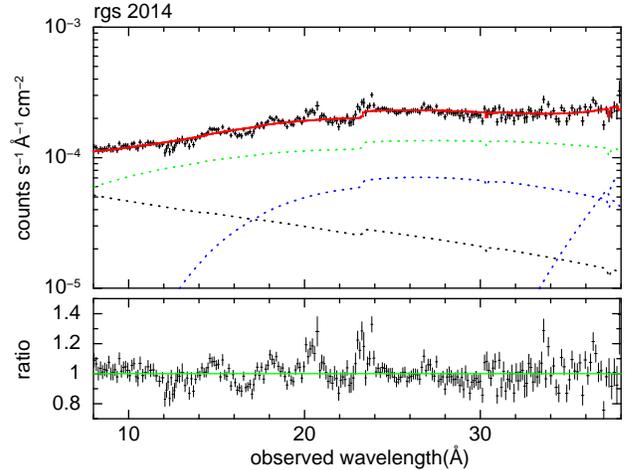}                                                                                         
\caption                                                                
{Continuum model consisting of hard ($\Gamma$=1.67) and soft ($\Gamma$=2.9) power law components (black and green, respectively) taken from the higher energy spectral fit reported in Paper 1, with the
addition of two blackbody components (blue) to match the 'soft excess'. The ratio of data to continuum illustrates the spectral structure imprinted on the x-ray continuum by
photoionized absorption and emission from an outflowing wind} 
\end{figure}

\subsection{Spectral modelling from 12--30 \AA\ with photoionized absorption and emission spectra}

We begin spectral modelling over the 12--30 \AA\ waveband where the data are of the highest statistical quality and the continuum fit gives a measure of spectral structure with \rchi\ of 834/452.
Publicly available photoionized grids computed for a power law $\Gamma$ = 2 ionizing continuum 
\footnote {ftp://legacy.gsfc.nasa.gov/software/plasma\_codes/xstar.xspectables} 
failed to fit strong Fe UTA absorption at $\sim$16--17 \AA\ without an
unrealistically high Fe abundance. Since absorption modelling can be highly sensitive  to the form of the photoionizing continuum (e.g. see Reeves et al. 2013),  we therefore generated a self-consistent
set  of multiplicative absorption and additive emission grids using the {\tt XSTAR} v2.2 code  (Kallman et al. 1996), with the spectral energy distribution (SED) of \pg\ based on concurrent data from the
Optical  Monitor (Mason \et 2001) and EPIC-pn cameras, extrapolated from 1-1000 Rydberg. The SED, described in more detail in Lobban \et (2016), is represented by a double-broken
power-law, where the OM data have been corrected for reddening (E(B-V) = 0.035).  

The new {\tt XSTAR} grids were computed in 10000 spectral bins between 0.1-20 keV, with an ionizing luminosity (1--1000 Ryd)
of  3.8$\times 10^{45}$ erg s$^{-1}$ based on the SED fit, and over a range in ionization log $\xi$ = 0-4 (where $\xi$ = L$_{ion}$/n r$^{2}$) and column density $N_{\rm H}$= $10^{19} -
10^{22}$ cm$^{-2}$.  Solar abundances were adopted for abundant elements (Grevesse \& Sauval 1998), with additional grids computed for 3- and 5-fold over abundances of Fe.  In each case grids were
generated with turbulence velocities of $\sigma$ = 300, 1000 and 3000 km s$^{-1}$, and for each multiplicative absorption grid (mtable) we generated a corresponding emission grid (atable).

Absorption grids with a modest over-abundance of Fe were found to give the best fit to the mean 2014 RGS spectrum, and the present analysis is based on a  3-times-solar Fe abundance, with a turbulent velocity of 300 km s$^{-1}$ for
the absorber and 1000 and 3000 km s$^{-1}$ for low- and high-ionization photoionized emission, respectively, to encompass velocity broadening.
Free parameters in the absorption spectra are column density, ionization  parameter, and outflow (or inflow) velocity in the AGN rest frame. For the emission spectra the column density is fixed at a suitably low 
value ($N_{\rm H}$ = 10$^{20}$ cm$^{-2}$)to avoid significant opacity effects, with the ionization, flux level and velocity as free parameters.

The full {\tt xspec} model is TBabs*((po1 + bbody)*mt + at + po2), where components 'mt' and 'at' represent the mtable and atable grids of pre-computed photoionized absorption and emission spectra. Po2 represents the
unabsorbed power law and TBabs the Galactic absorption. The
absorbed power law has photon index ($\Gamma$=1.67) tied to the value found in the pn analysis, with a single blackbody  of kT$\sim$0.1 keV completing the soft excess at 12--30 \AA. 

The addition of a single photoionized absorber greatly improved the 12-30 \AA\ spectral fit (\rchi\ of 681/451), for an ionization parameter log $\xi$=1.64$\pm$0.06, absorption column density  $N_{\rm
H}$ = 5.0$\pm$0.6$\times 10^{20}$ cm$^{-2}$ and outflow velocity 0.061$\pm$0.001c, close to the lower of the dual velocities reported in Paper 1.  Adding a photoionized emission spectrum, with tied ionization
parameter, produced a further substantial improvement to the fit (\rchi\ of
601/449), with the linked ionization parameter increasing to log $\xi$=1.77 $\pm$0.05, and the absorber  column falling slightly to $N_{\rm H}$ = 4.3$\pm$0.6$\times 10^{20}$ cm$^{-2}$. The absorber
outflow velocity was unchanged, while the integrated emission spectrum had a much lower outflow velocity in the AGN rest frame.

The interaction of emission and absorption grids can be partly understood by the effects of self absorption in resonance lines, while correctly modelling emission spectra also ensures a more accurate
fit to the intrinsic continuum. Joint emission and absorption modelling also played a key role in analysis of the complex Fe K absorption spectrum reported in Paper 1, and underlines the importance of including both emission and
absorption  spectra in outflow studies of high quality data. 

The addition of a second ionized absorber gave a further substantial improvement to the 12--30 \AA\ spectral fit (\rchi\ of 569/446), with an ionization parameter log
$\xi$=1.8$\pm$0.1, absorption  column density ($N_{\rm H}$ = 2.1$\pm$0.5$\times 10^{20}$ cm$^{-2}$) and a significantly higher outflow velocity of 0.077$\pm$0.001c. A
second emission spectrum with tied ionization parameter yielded only a small further improvement in the spectral fit (\rchi\ of 561/444).

Addition of the second absorber had the ancillary effect of reducing the ionization parameter of the  first - lower velocity -  absorber to log $\xi$=1.33$\pm$0.08, with a further decrease in
column density  to $N_{\rm H}$ = 3.9$\pm$0.6$\times 10^{20}$ cm$^{-2}$. While a second tied emission component had little effect, de-coupling the ionization parameters of both emission and absorption spectra 
did yield a further significant 
improvement in the fit (\rchi\ of 534/442), with the emission components now covering a wider range
of ionization. Table 1 summarises the parameters of this best-fit 12--30 \AA\ model, including the  luminosities added/extracted 
by the respective  ionized emission/absorption spectra. 

\begin{figure}                                                          
\centering                                                                                                                          
\includegraphics[width=6.2cm, angle=270]{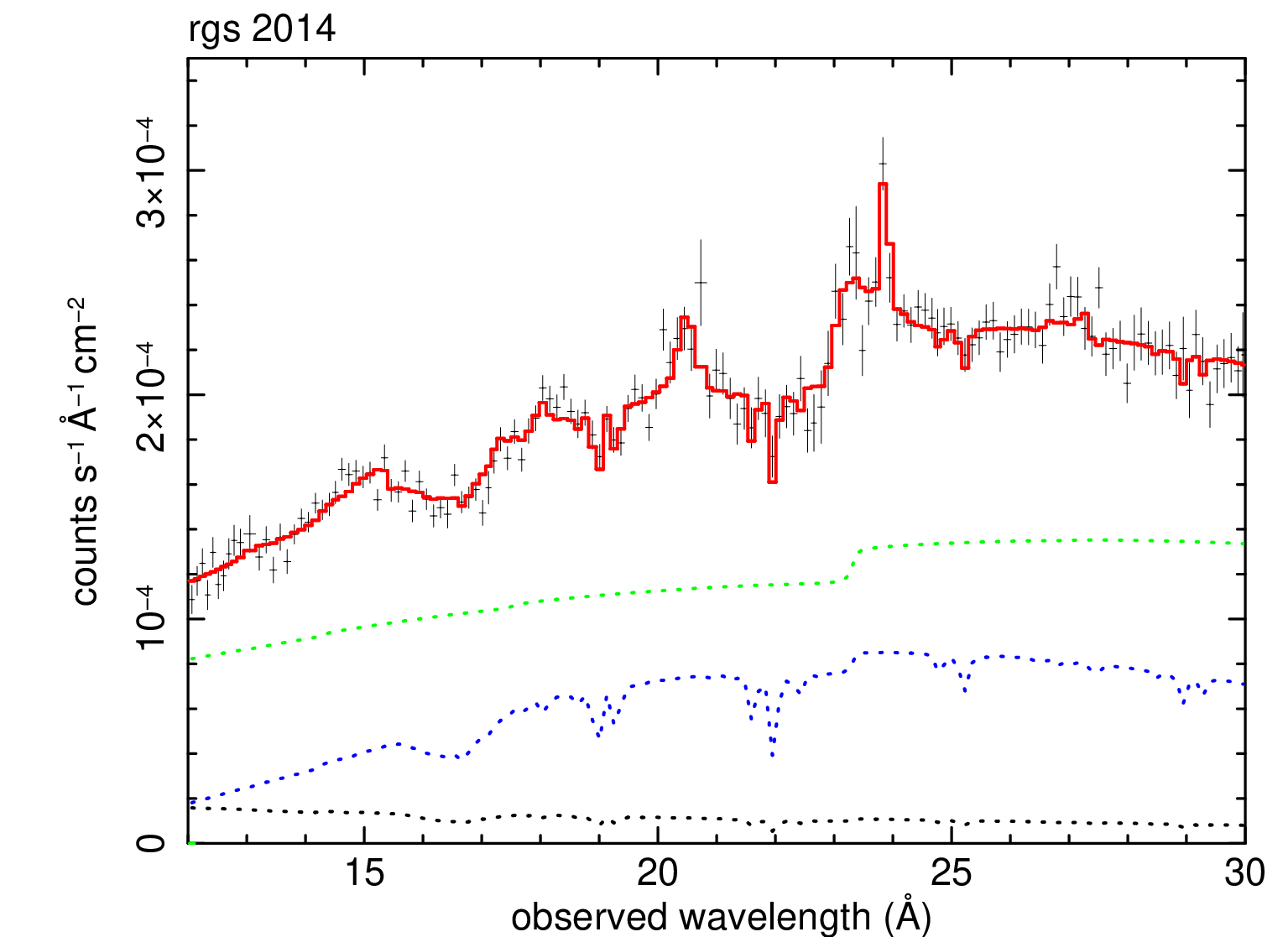}                                                
\centering                                                              
\includegraphics[width=6.18cm, angle=270]{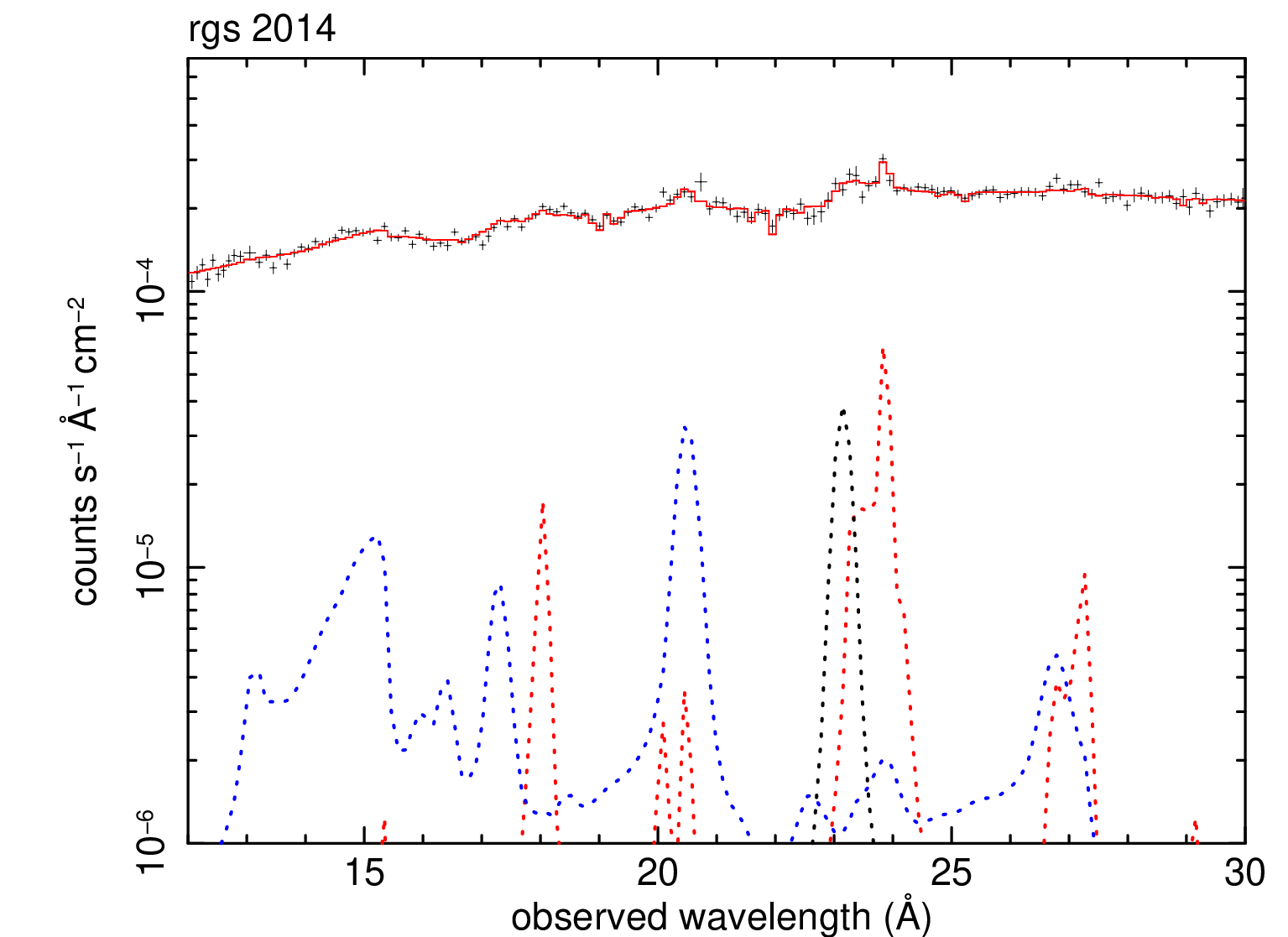}                                           
\centering                                                              
\includegraphics[width=5.85cm, angle=270]{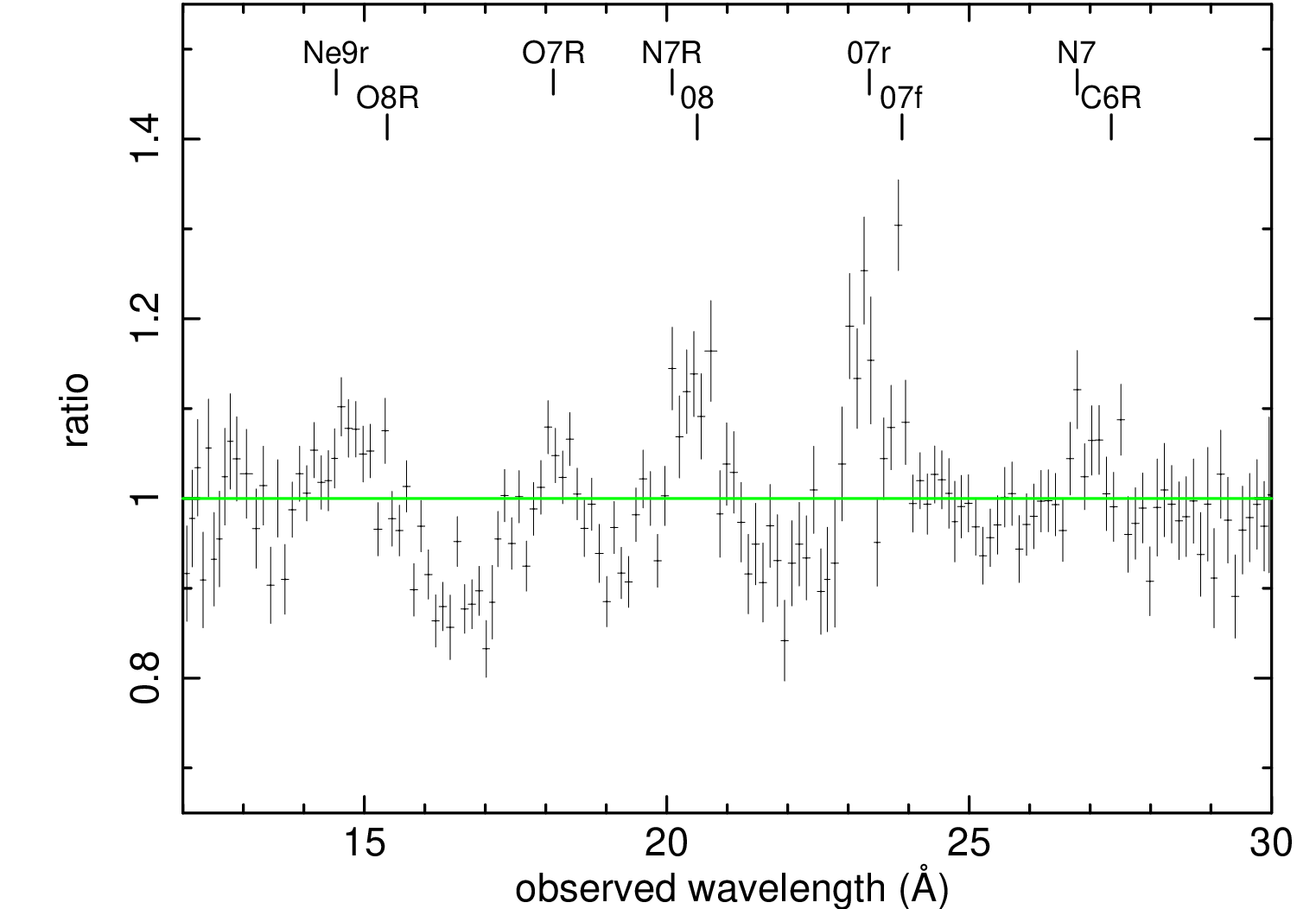}    
\caption                                                                
{(top) A section of the mean soft X-ray spectrum of \pg\ from the stacked RGS data of the 2014 \xmm\ observation, with photoionized absorption and emission added by {\tt XSTAR} grids as described in the text. In the
model the absorption affects only the hard power law (black) and  blackbody (blue) continuum but not the steeper power law (green) identified in inter-orbit difference spectra. (mid panel) Emission
spectra from low (red) and high (blue) ionization components, with the parameters listed in Table 1. (lower) RGS data plotted as a ratio to the underlying continuum, with several discrete emission
features identified with resonance emission lines and radiative recombination continua (R) of H- and He-like ions of Ne, O, N and C.}      
\end{figure}  
  
Figure 2 illustrates different aspects of the 12--30 \AA\ spectral fit.  The top panel compares the data and final model, together with the three continuum components, unabsorbed power law (green), hard
power law (dark blue) and black body (light blue). Absorption over most of the RGS waveband is seen to predominently affect the black body emission component, representing the 'soft excess', a point which may be
important in comparing the column density of similar flows measured in soft and hard x-ray spctra (see Section 6).

Strong absorption features include the Fe-M  UTA at
$\sim$16--17 \AA\ (Behar \et\ 2001), and resonance absorption in OVIII Lyman-$\alpha$ and the OVII triplet observed at $\sim$19 \AA\ and $\sim$22 \AA, respectively. We find in Section 4 that the twin lines
apparent in each resonance transition correspond to the distinct  velocities found in the {\tt xstar} modelling (Table 1). In turn, both low velocity outflow components contribute to the strong Fe UTA, with
a fit limited to the 14--18 \AA\ waveband confirming both velocities are required (see also Appendix). We note this finding differs from that in the deep \chandra\ observation of NGC 3783, 
where Holczer \et (2005) found the Fe
M-shell absorber to be separated from the main outflow and essentially stationary.

\begin{table*}
\centering
\caption{Parameters of the ionized outflow from 12--30 \AA\ fit to the combined 2014 RGS data, with two photoionized absorbers, each defined by an ionization parameter  $\xi$ (erg cm s$^{-1}$), column
density $N_{\rm H}$ (cm$^{-2}$),  outflow velocity (in units of c) and absorbed luminosity (erg s$^{-1}$), and two photoionized emission spectra defined by the respective ionization parameter, outflow
velocity, normalisation and luminosity.  Velocities are in the AGN rest frame and  luminosities are in the fitted spectral band. The significance of each spectral component is measured by the increase in $\chi^{2}$ 
when that component is removed and the spectrum re-fitted. }
\begin{tabular}{@{}lccccc@{}}
\hline
component & log$\xi$ & $N_{\rm H}$ ($10^{20}$)  & v$_{out}/c $ & L$_{abs/em}$ & $\Delta \chi^{2}$ \\
\hline
absorber 1 & 1.35$\pm$0.09 & 3.6$\pm$0.7 & 0.061$\pm$0.001  & 1.5$\times10^{42}$ & 60/3 \\
absorber 2 & 1.8$\pm$0.1 & 1.9$\pm$0.5  & 0.077$\pm$0.001  & 6$\times10^{41}$ & 23/3 \\
emitter 1 & 1.1$\pm$0.3 & 10 & 0.0016$\pm$0.0005 & 6$\times10^{41}$ & 59/3 \\
emitter 2 & 3.1$\pm$0.3 & 10 & 0.001$\pm$0.002 & 1.0$\times10^{42}$ & 33/3 \\
\hline
\end{tabular}
\end{table*}

The modelled photoionized emission spectra are highlighted in the mid panel of Figure 2, and directly compared with the stacked data in the lower panel,
where the principal emission  features are identified with resonance emission lines and  radiative recombination continua (RRC) of the abundant H- and He-like ions of Ne, O, N and C.  A more detailed 
comparison of data and model identifies a relative weakness of the modelled resonance emission in the triplet of OVII, suggesting a missing contribution from photoexcitation (Kinkhabwala \et\
2002) not included in the {\tt xstar} grids. Adding a  Gaussian emission line to the model described above finds a (blue-shifted) line energy of 0.536$\pm$0.002 keV (23.13 \AA) and
equivalent width 1.6$\pm$0.4 eV, yielding an improvement in $\Delta \chi^{2}$ of  22/3.

In summary of Section 2, modelling of spectral structure in the 12--30 \AA\ waveband finds strong evidence  for absorption in a photoionized outflow, with velocity components $\sim$0.061c and
$\sim$0.077c, which we assume are blended to match the lower of the dual velocities (v$\sim$0.066c) identified in the primary highly ionized outflow in the (lower resolution) hard x-ray data (Paper 1).  

From the model spectral fit we obtain an overall spectral luminosity in the 12-30 \AA\ waveband of 5.73$\times 10^{43}$ erg s$^{-1}$, with the low ionization absorbers together removing 2.1$\times 10^{42}$ erg
s$^{-1}$. The low ionization emission spectrum contributes a luminosity over the same waveband of 6$\times 10^{41}$ erg s$^{-1}$, suggesting a covering factor of that
flow component of $\sim$30 \%.  We return  to considerations of the energy budget in Section 5.

\section{Extending the RGS spectral fit from 8-38 \AA\ waveband}

While the lower of the two outflow velocities (v$\sim$0.066c) reported in Paper 1 may be clearly identified, the higher velocity
(v$\sim$0.129c) seen in the pn data is not detected. Since that difference might be due the faster flow being more highly ionized (as was
found in the  pn data of Paper 1), and the soft x-ray spectral model developed in Section 2 includes emission but not absorption from highly ionised outflow components,we now extend spectral modelling 
over the full 8--38 \AA\ waveband shown
in Figure 1.

The more extended spectral fit confirms the two absorption
components (absorber 1 and absorber 2) listed in Table 1, which are now included with only
minor changes in the relevant parameters (abs1 and abs2) in Table 2. 
Significantly, the increased high energy coverage now also finds highly ionized counterparts of both high velocity outflow components reported in Paper 1. 
Visual examination of the new spectral model confirms these new detections (abs3
and abs4 in Table 2) are largely due to several strong absorption lines of Fe XVIII -- XX and resonance lines of Ne X
and Mg XI observed in the  8--14 \AA\ spectral region. The lower panels of Fig.3 illustrate this point.

\begin{figure}                                                                                                  
\centering 
\includegraphics[width=6.3cm, angle=270]{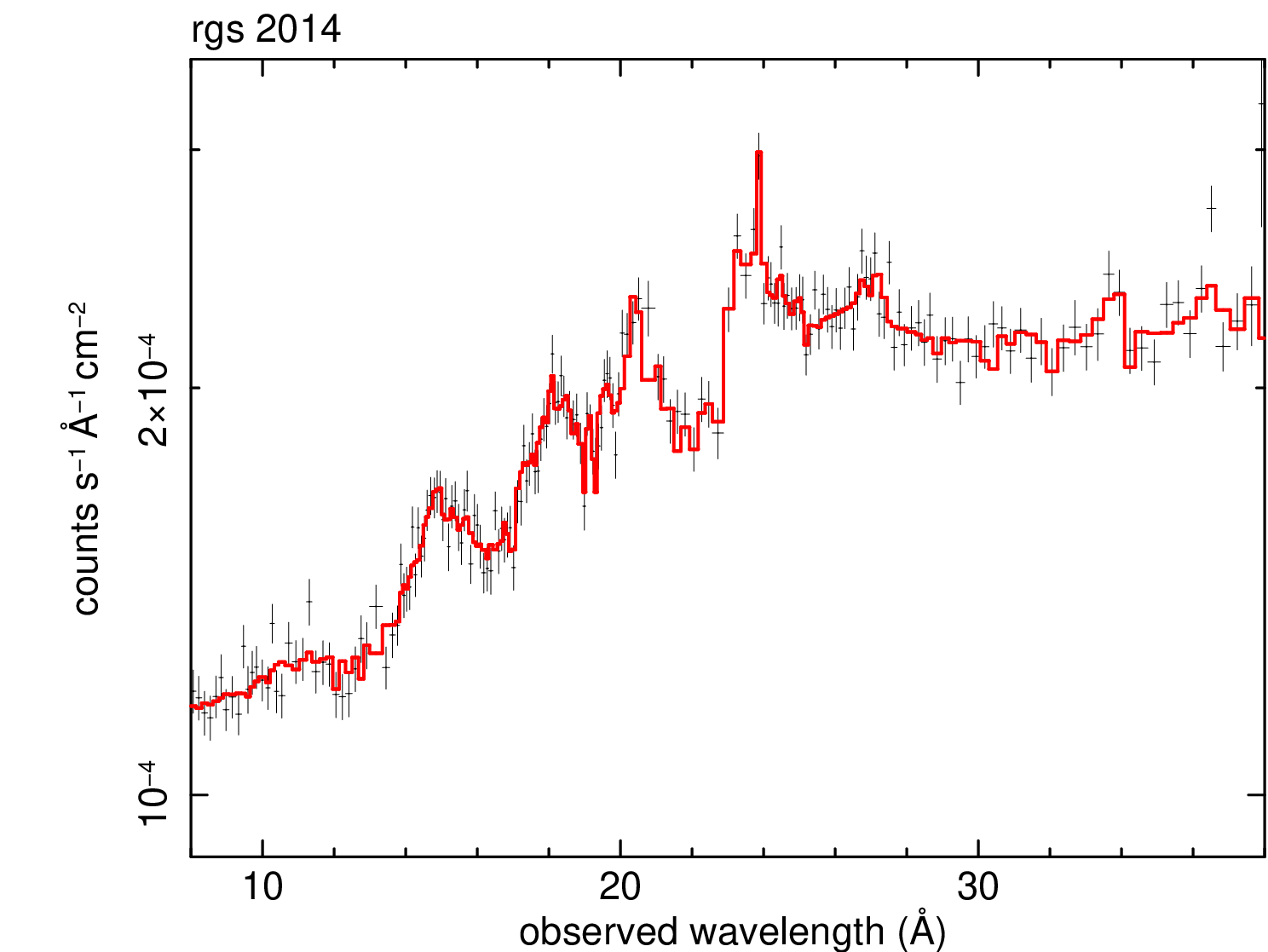}                                                                                                                                                                                 
\centering 
\includegraphics[width=3.15cm, angle=270]{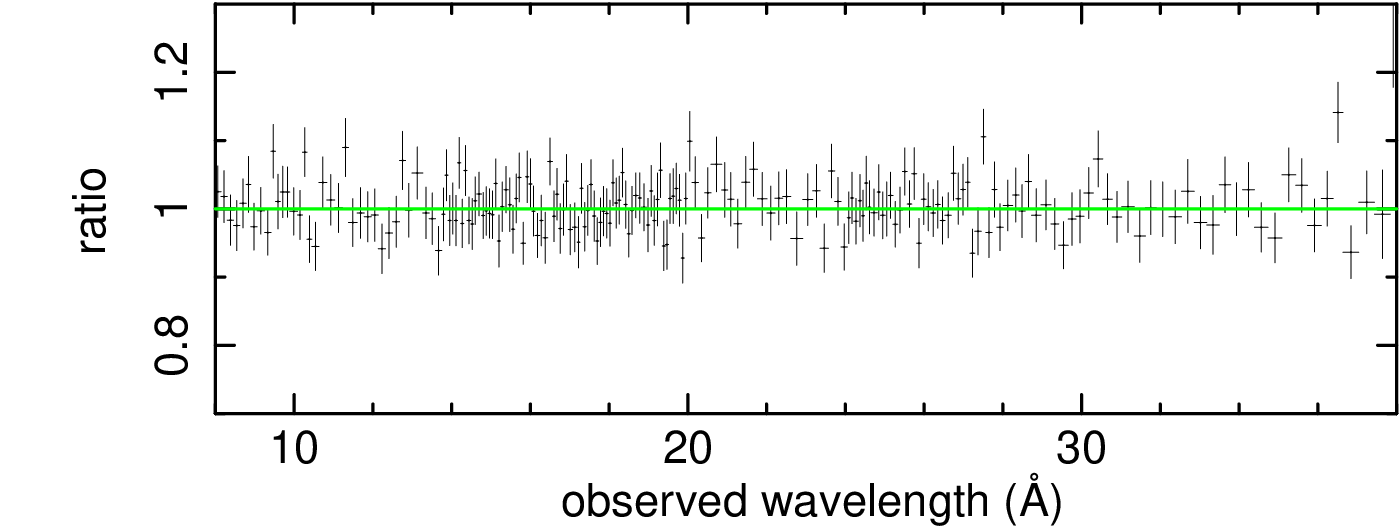}                                                                                                                                                                                 
\centering 
\includegraphics[width=2.77cm, angle=270]{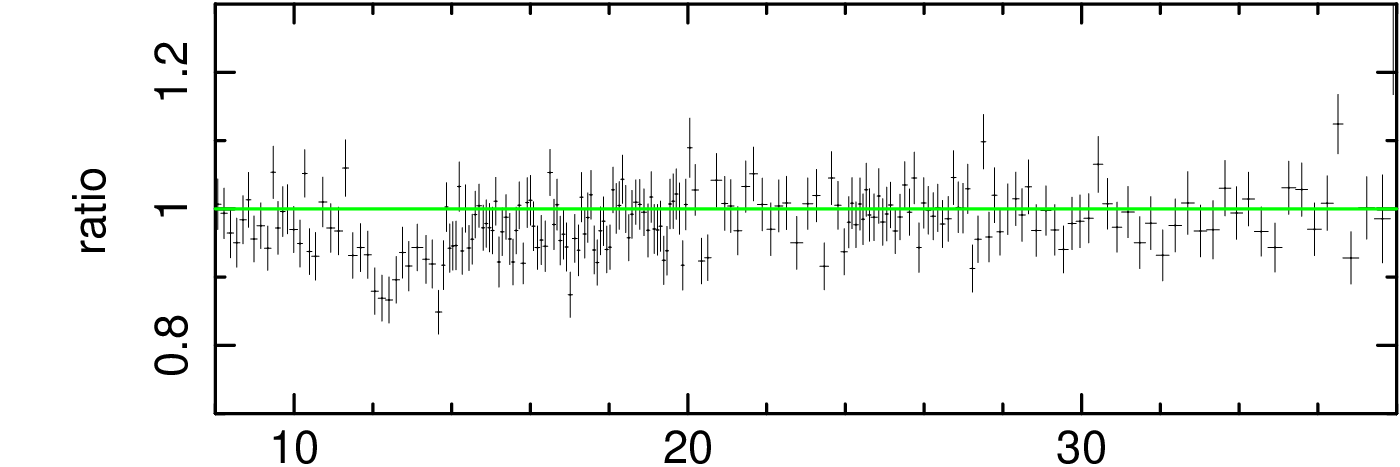}          
\caption                                                                
{(top) The soft x-ray spectrum of \pg, where fine structure on the soft x-ray continuum in Figure 1 is successfully modelled by photoionized absorption and emission spectra from an outflowing wind
with the parameters listed in Table 2. The middle panel shows the data-model residuals of this multi-component fit, where comparison with the lower panel - based on the restricted model of Table 1 - provides
visual confirmation of the contributions of highly ionized components abs3 and abs 4 in the extended 8--38 \AA\ spectral fit} 
\end{figure}

The extended model of the RGS soft x-ray spectrum now includes both low and highly ionized absorbers with
velocities close to those identified in the hard x-ray spectrum reported in Paper 1. In addition to offering compelling support for both
those primary (high column) outflow components, the higher resolution grating spectra provide an indication of additional fine structure
in the lower velocity flow. Modelling the ionized emission with components of widely differing ionization parameter is a further measure
of a complex flow.

A final search for additional absorption components to those noted above, provides evidence for a still higher velocity outflow of 0.188$\pm$0.001c (abs5 in Table), given added weight as that velocity coincides 
closely with a 
marginal detection of a third highly ionized outflow component in the hard x-ray spectrum reported in Paper 1 (see also Appendix).

\begin{table*}
\centering
\caption{Parameters of the ionized outflow of \pg\ over the full 8--38 \AA\ waveband. Five photoionized absorbers are detected, each defined by an ionisation parameter  $\xi$ (erg cm s$^{-1}$), column
density $N_{\rm H}$ (cm$^{-2}$),  outflow velocity (km s$^{-1}$) and absorbed luminosity (erg s$^{-1}$), with two photoionized emission spectra defined by the respective ionization parameter, outflow
velocity, normalisation and luminosity.  Velocities are in the AGN rest frame and luminosities relate to the fitted spectral band.}
\begin{tabular}{@{}lccccc@{}}
\hline
comp & log$\xi$ & $N_{\rm H}$ ($10^{20}$) & v$_{out}$/c & L$_{abs/em}$ & $\Delta \chi^{2}$ \\
\hline
abs 1 & 1.5$\pm$0.1 & 2.3$\pm$0.4 & 0.061$\pm$0.001  & 1.6$\times10^{42}$ & 60/3  \\
abs 2 & 1.9$\pm$0.2 & 1.5$\pm$0.5  &  0.077$\pm$0.001  & 7$\times10^{41}$ & 36/3  \\
abs 3 & 3.0$\pm$0.2 & 13$\pm$6 &  0.131$\pm$0.001 & 6$\times10^{41}$ &  15/3   \\
abs 4 & 3.2$\pm$0.1 & 28$\pm$15 & 0.061$\pm$0.001   & 8$\times10^{41}$ &  18/3 \\
abs 5 & 2.5$\pm$0.1 & 12$\pm$5 & 0.188$\pm$0.002   & 9$\times10^{41}$ &  18/3 \\
emit 1 & 1.7$\pm$0.3 & 10 & 0.0012$\pm$0.0006  & 9$\times10^{41}$ &  73/3  \\
emit 2 & 2.8$\pm$0.6 & 10 & 0.002$\pm$0.002  & 3$\times10^{41}$ &  13/3  \\
\hline
\end{tabular}
\end{table*}

Based on the extended spectral fit, the observed spectral luminosity from 8-38 \AA\ is 9.2$\times 10^{43}$ erg s$^{-1}$, with absorption components 1 and 2, corresponding to the lower of the two velocities 
found in Paper 1, having removed 
2.3$\times 10^{42}$ erg s$^{-1}$, with components 3 an 4, associated with the more highly ionized outflow, removing 2.4$\times 10^{42}$ erg s$^{-1}$. A lower ratio 
of emission-to-absorption  for the highly ionized components may indicate a lower covering factor, or 
that much of the re-emission lies outside the RGS spectral band. We return to that question in Section 5.

Figure 3 overlays the 2014 RGS data with the final spectral model. Comparison of the data-model residuals with those of Fig. 1
illustrates how the addition of the photoionized absorption and emission spectra detailed in Table 2 greatly improve the overall
spectral fit.

\section{Identifying individual resonance lines}

As noted earlier, the high ionization characteristic of powerful AGN winds renders much of that outflow transparent at soft x-ray energies, while the presence of an unabsorbed continuum component has a further
diluting  effect on observed soft x-ray absorption features.  An added complexity in the soft x-ray spectrum of \pg\ is the dominance across parts of the RGS waveband of multiple transitions in Fe-L,
including those partially screened by M-shell electrons  (the Fe-M UTA, Behar \et\ 2001). While the complexity of the Fe-L absorption can be accounted for reasonably well in spectral modelling, 
it is difficult to reliably identify individual transitions for use in measuring a blue-shift.

Fortunately, the unusually deep \xmm\ observation in 2014 has allowed the detection and identification of several absorption lines with physically resonance transitions in Ne, O, N and C (Figure 4). 
Sequentially stepping through the stacked data with narrow Gaussians of width tied to the mid-band RGS resolution ($\sim$30 m\AA) yields the observed absorption line wavelengths listed in Table 3, where each
line identification provides a corresponding blue-shift and outflow velocity. The 90\% confidence error on each wavelength was obtained using the {\tt uncertainty} command for the relevant Gaussian
parameter, which was then carried over to give the listed velocity uncertainty.

\begin{figure*}                                                                                                  
\centering 
\includegraphics[width=5.8cm, angle=270]{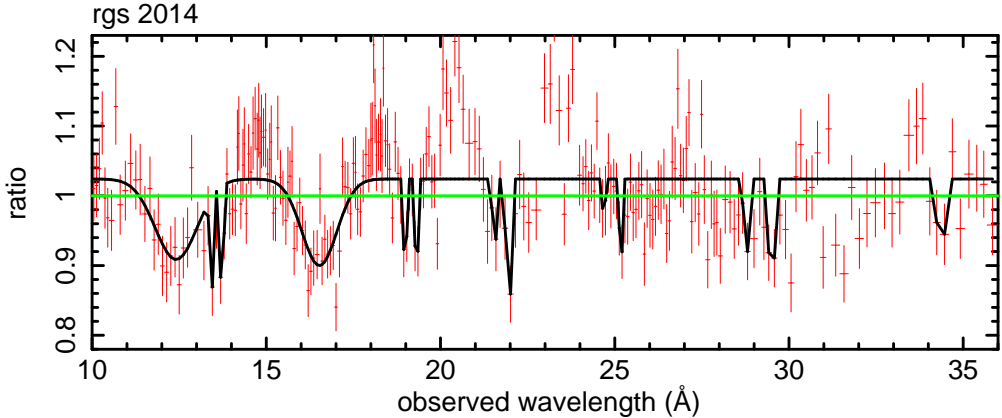}                                                                                         
\caption                                                                
{Gaussian line fitting to spectral structure in the stacked soft x-ray spectrum of \pg\ when the absorption and emission spectra defined in Table 2 have been removed from the model.  A blind search
reveals the 11 narrow  absorption lines listed in Table 3, with 5 and 6 lines, respectively, found to match the two low ionization outflow velocities obtained from spectral modelling, when identified
with principal resonance lines of Ne, O and N. Two broader Gaussians, at $\sim$ 12--13 \AA\ and $\sim$ 16--17.5 \AA\ are used to fit a complex group of strong Fe L lines and the Fe-M UTA,
respectively.}  
\end{figure*}

\begin{table*}
\centering
\caption{Observed wavelength of narrow Gaussians fitted to apparent absorption lines in the stacked 2014 RGS spectrum. 11 narrow lines are identified with
six resonance transitions in Ne, O, N and C, with common blue-shifts indicating two outflow velocities, both consistent with values obtained from spectral modelling}
\begin{tabular}{@{}lcccccc@{}}
\hline
line & lab (\AA) & obs (\AA) & rest (\AA) & v$_{out}$/c & $\Delta \chi^{2}$ \\
\hline
Ne9 He-$\alpha$ & 13.447 & 13.41$\pm$0.03 & 12.41$\pm$0.03 & 0.080$\pm$0.002 & 10/2 \\
Ne9 He-$\alpha$ & 13.447 & 13.73$\pm$0.02 & 12.70$\pm$0.02 & 0.057$\pm$0.002 & 15/2 \\
O8 L-$\alpha$ & 18.968 & 18.99$\pm$0.02 & 17.57$\pm$0.03 & 0.077$\pm$0.002 & 12/2 \\
O8 L-$\alpha$ & 18.968 & 19.31$\pm$0.02 & 17.86$\pm$0.03 & 0.060$\pm$0.002 & 16/2 \\
O7 He-$\alpha$ & 21.602 & 21.54$\pm$0.04 & 19.93$\pm$0.03 & 0.080$\pm$0.003 & 4/2 \\
O7 He-$\alpha$ & 21.602 & 21.94$\pm$0.03 & 20.29$\pm$0.03 & 0.062$\pm$0.002 & 20/2 \\
N7 L-$\alpha$ & 24.781 & 24.69$\pm$0.05 & 22.84$\pm$0.03 & 0.080$\pm$0.003 &  4/2 \\
N7 L-$\alpha$ & 24.781 & 25.17$\pm$0.02 & 23.28$\pm$0.03 & 0.062$\pm$0.002 &  11/2 \\
N6 He-$\alpha$ & 28.787 & 28.75$\pm$0.04 & 26.60$\pm$0.03 & 0.078$\pm$0.003 &  7/2 \\
N6 He-$\alpha$ & 28.787 & 29.50$\pm$0.05 & 27.29$\pm$0.03 & 0.054$\pm$0.003 &  14/2 \\
C6 L-$\alpha$ & 33.736 & 34.35$\pm$0.15 & 31.78$\pm$0.03 & 0.060$\pm$0.004 &  5/2 \\
\hline
\end{tabular}
\end{table*}

For the 9 lines with $\Delta \chi^{2}$ of 5 or greater, two distinct velocity groups are found, with weighted mean outflows of v$\sim$0.060$\pm$0.002c and v$\sim$0.078$\pm$0.002c, in close agreement with
the velocities of the two soft x-ray absorbers found in the {\tt XSTAR} modelling. That agreement provides strong, independent support for those modelled outflow velocities, and correspondingly for the similar low
velocity component found in the primary, highly ionized outflow of \pg\ reported in Paper 1. The absence of identified absorption lines from the higher velocity outflow (v$\sim$0.13c) may be understood
from the relatively high ionization parameter of that component (abs3 in Table 2), with the strongest absorption falling in the 12--14 \AA\ waveband dominated by Fe-L transitions. This absorption complex is
represented by a broad Gaussian in Figure 4, where only two narrow lines are identified, both with NeIX He-$\alpha$. Interestingly, one narrow absorption line at $\sim$17 \AA\ does stand out from the broad Gaussian 
fit to the Fe UTA
and we see in the Appendix that this very likely corresponds to OVIII Ly-$\alpha$ outflowing at the highest velocity component (abs5) in Table 3.

\section{Comparing the soft and hard x-ray spectral models}

The above analysis has shown that the dual outflow velocities, of $\sim$0.129c and $\sim$0.066c, identified in highly ionized absorbing matter in the 2014 pn data (Paper 1) have co-moving counterparts
in the soft x-ray RGS spectra, with the higher spectral resolution of the grating spectra suggesting the lower velocity is actually a blend of two components.    

As in Paper 1, simultaneous modelling of photoionized emission spectra is found to be important, both in determining the correct absorption parameters of the flow and in showing that a substantial fraction
of the absorbed energy is returned via  recombination and scattering from an extended outflow. To evaluate the energy budget of this re-processing requires extending the spectral bandwidth beyond that
covered by the RGS, not least as the major part of the flow is absorbed only in the higher energy band covered by the EPIC cameras.  

A direct comparison with the spectral model obtained from analysis of the simultaneous pn data in Paper 1 shows a good measure of agreement - and some important differences.  To that end, the 2--10 keV spectral fit 
from Paper 1 was  extended down to 0.4 keV to overlap the RGS spectrum. In making that
extension the soft x-ray continuum was modelled by the addition of a black body component (kT$\sim$0.1 keV), with soft x-ray absorption and emission requiring additional emission and absorption grids
compared with Paper 1. 

The extended 0.4--10 keV pn spectral fit remained remarkably good (\rchi\ of 1783/1676), with the emission and absorption parameters listed in Table 4. 

Comparison of the low ionization absorbers in Tables 2 and 4 tests the conjecture that abs3 in Table 4 encompasses both abs1 and abs2 absorbers in the RGS spectra, with the
velocity of v$\sim$0.066c being a blend of two outflow components unresolved in the CCD data. Similar absorbed luminosities and column densities of the low ionization absorbers
in the RGS and extended pn spectral models are now seen to support that view.  

More surprising is the large difference in the column densities of highly ionized absorbers with the same velocity, but measured against the soft x-ray continuum (Table 2, abs3 and abs4) as compared with
the hard x-rey
spectrum (Table 4, abs1 and abs2). In Section 6 we suggest an explanation in terms of different sight lines through the same outflow component, with potential to explore the x-ray source geometry.

\begin{table*}
\centering
\caption{Parameters of the ionized outflow from the previously published 2--10 keV spectral fit, extended down to 0.4 keV to overlap the soft x-ray spectrum covered by the 
RGS data. The continuum is modelled with absorption of a hard power law $\Gamma$ $\sim$1.67 and blackbody (kT $\sim$0.1 keV), together with a softer 
unabsorbed power law of photon index $\Gamma$ $\sim$2.9. Velocities are in the AGN rest-frame and luminosities in the 0.4--10 kev spectral band}
\begin{tabular}{@{}lccccc@{}}
\hline
comp & log$\xi$ & $N_{\rm H}$ ($10^{22}$)  & v$_{out}$/c & L$_{abs/em}$ & $\Delta \chi^{2}$ \\
\hline
abs 1 & 3.95$\pm$0.25 & 36$\pm$24 & 0.129$\pm$0.002  & 6$\times10^{41}$ & 19/3 \\
abs 2 & 3.50$\pm$0.07 & 20$\pm$9  & 0.066$\pm$0.002  & 1.2$\times10^{42}$ & 41/3 \\
abs 3 & 1.7$\pm$0.2 & 0.05$\pm$0.02 & 0.066 (t)  & 1.4$\times10^{42}$ & 24/3 \\
emit 1 & 3.47$\pm$0.05 & 1 & 0.007$\pm$0.004 & 1$\times10^{42}$ & 22/2 \\
emit 2 & 1.46$\pm$0.07 & 0.1 & 0.002$\pm$0.002 & 1.3$\times10^{42}$ & 15/2 \\
\hline
\end{tabular}
\end{table*}

An indication of the large covering factor of the outflow in \pg\ is now best measured from the broad band 0.4-10 keV fit, where we find a total luminosity of 1.47$\times 10^{44}$ erg s$^{-1}$, with the
high and  low ionization photoionized absorbers removing 1.8$\times 10^{42}$ erg s$^{-1}$ and 1.4$\times 10^{42}$ erg s$^{-1}$, respectively, or $\sim$2.2\% of the spectral luminosity. In comparison, the
high and low ionization emission spectra contribute  luminosities over the same spectral band of 1$\times 10^{42}$ erg s$^{-1}$ and 1.3$\times 10^{42}$ erg s$^{-1}$, respectively, with the total ionized emission
indicating a time-averaged covering factor of $\sim$60\%.

\section{Discussion}

An analysis of the soft x-ray spectrum of \pg, produced by stacking RGS spectral data from all seven orbits of the 2014 \xmm\ observation, has revealed both low and highly ionization outflow
components co-moving with each of the two primary high velocity outflows detected in the hard x-ray spectrum (Paper 1). The higher resolution of the grating spectrometers reveals additional velocity structure, with
outflow velocities of 0.061$\pm$0.002c and 0.077$\pm$0.002c, apparently blended in the $\sim$0.066c component detected in the CCD spectra.

As in Paper 1 we find photoionized emission spectra are of similar significance to photoionized absorption in modifying the observed soft x-ray spectrum, with {\tt xstar} modelling successfully  reproducing
most of the observed features, dominated by resonance emission lines and radiative recombination continua (RRC) of the abundant lighter metals (C, N, O, Ne) having K-shell energies covered by the RGS. 

Extending the 2-10 keV spectral fit of Paper 1 down to 0.4 keV provides a  direct comparison of the hard and soft x-ray analyses, allowing a quantitative measure of the impact of the photoionized outflow on the
observed x-ray spectrum. From that assessment  we find the outflow  removes $\sim$2.3\% of the 0.4--10 keV spectral luminosity, with the ionized emission returning $\sim$80\% of that luminosity 
from scattering
and recombination in an apparently extended outflow. 

The relative column densities of co-moving low and high ionization absorption components in Table 2 suggest an outflow in line of sight to the soft x-ray continuum source with a small filling 
factor of embedded higher density matter. Adopting that explanation, comparison of abs1 and abs4 (Table 2) shows a factor $\sim$12 in column
density and $\sim$50 in particle density, indicating a linear filling factor of $\sim$ 0.16\%. 
It appears that while such co-moving absorbers may be readily understood in terms of higher density 'clouds' embedded in the main flow, perhaps caused by instabilities in the flow, the small filling factor implies a 
negligible contribution to the outflow mass rate and momentum.

Explaining the large difference in column density for flow components  detected in the
RGS and pn data {\it and having the same velocity and ionization parameter} is less straightforward. The highly ionized outflow components detected in the hard x-ray data (abs 1 and abs 2 in Table 4) have column
densities $\sim$300 and $\sim$70 times larger than their soft x-ray counterparts (abs3 and abs4, respectively, in Table 2). We suggest an explanation might involve different lines-of-sight to the dominant hard
and soft x-ray continuum sources (power law and blackbody in the model; Figure 1). 

For a conventional disc-corona geometry the intrinsic hard x-ray source would probably be confined to a smaller radius, closer to the black hole and - importantly - to the likely wind launch site (King and Pounds
2003), compared with the more extended thermal soft x-ray source in the inner accretion disc. For a sufficiently small launch radius, a diverging wind (King and Pounds 2015) might then explain the large difference in
column density of the same flow component, observed in different sight lines to the hard and soft x-ray continuum sources. If confirmed, that possibility would offer the intriguing potential of using future 
observations of outflow spectra to
explore the geometry of the x-ray continuum source in an AGN.

Finding multiple UFO velocities represents a new challenge to models of powerful AGN winds, not least to continuum-driving (King and Pounds 2003) which provides a natural mechanism for the highly ionized matter 
characteristic of UFOs, and has been found to
provide a good match to their generic properties (King and Pounds 2015).
We noted in Paper 1 that chaotic accretion (King and Pringle 2006),  consisting of random
prograde and retrograde events, could offer an explanation of a dual velocity wind, the two distinct outflow velocities relating to different orientations of the current inner accretion flow. With both
flows close to Eddington, the prograde and retrograde discs would have different limiting values of the accretion efficiency $\eta$ and hence of velocity (King and Pounds 2003). 
Confirmation here of a third outflow velocity, marginally detected in Paper 1, may indicate further complexity in the inner accretion flow, which remains the natural site for the launch of a high speed,large
column wind. A further constraint is the relative variability in the primary (high colunn) flow components over the \xmm\ observations of 2001, 2007 and 2014 which will be reported in a future paper.

Meanwhile, in summary, the deep RGS exposure of \pg\ has provided an impressive demonstration of the application of high resolution spectra in studying the properties of a classical UFO. However, while the imprint of
overlying ionized gas on the soft x-ray spectrum provides detailed information on the dynamics of that outflow, with the soft x-ray absorption and emission being a high resolution tracer of lower
ionization matter entrained in the more massive highly ionized flows, it is important to note that the mass, momentum and mechanical energy rates of the UFO in \pg\  remain dominated by the much high
column  densities detected in the highly ionized (primary) flow components.

\section{Appendix}

We noted in Section 4 that it was difficult to identify the complex of Fe-L absorption lines at $\sim$ 11--14 \AA\ and in the strong Fe-M UTA at $\sim$ 15.5--17 \AA\
sufficiently reliably to allow a blue shift (and velocity) determination. That issue is made still more difficult when multiple velocities are involved, as illustrated in Figure 5.
There, the contributions of the v$\sim$0.061c and v$\sim$0.077c low ionization absorbers are shown separately, with the velocity difference resulting in a broader UTA
than for a single velocity. 

The upper panel of Fig.5 also provides a direct confimation of several individual line identifications, for example those at $\sim$ 19 \AA\ and $\sim$ 19.3 \AA\  identified with 
OVIII Ly-$\alpha$ and associated with the above outflow velocities in Table 3.
The lower panel of Fig. 5 shows the absorption profile for the highly ionized outflow component with v$\sim$0.188c (Table 2, abs 5), where the same OVIII Ly-$\alpha$ line is observed at $\sim$17 \AA.
While not fitted by a narrow Gaussian in Fig.4, for the reasons noted above, a strong absorption line is clearly visible at that wavelength. 

The important point to note is that while individual blue-shifted line identifications are an important means of identifying an outflow - and a co-moving line-set is better still - 
the most secure way to determine the parameters of any UFO is by spectral modelling, as described in Sections 2 and 3 for the present case of \pg.

\begin{figure}                                                                                                  
\centering 
\includegraphics[width=5.8cm, angle=270]{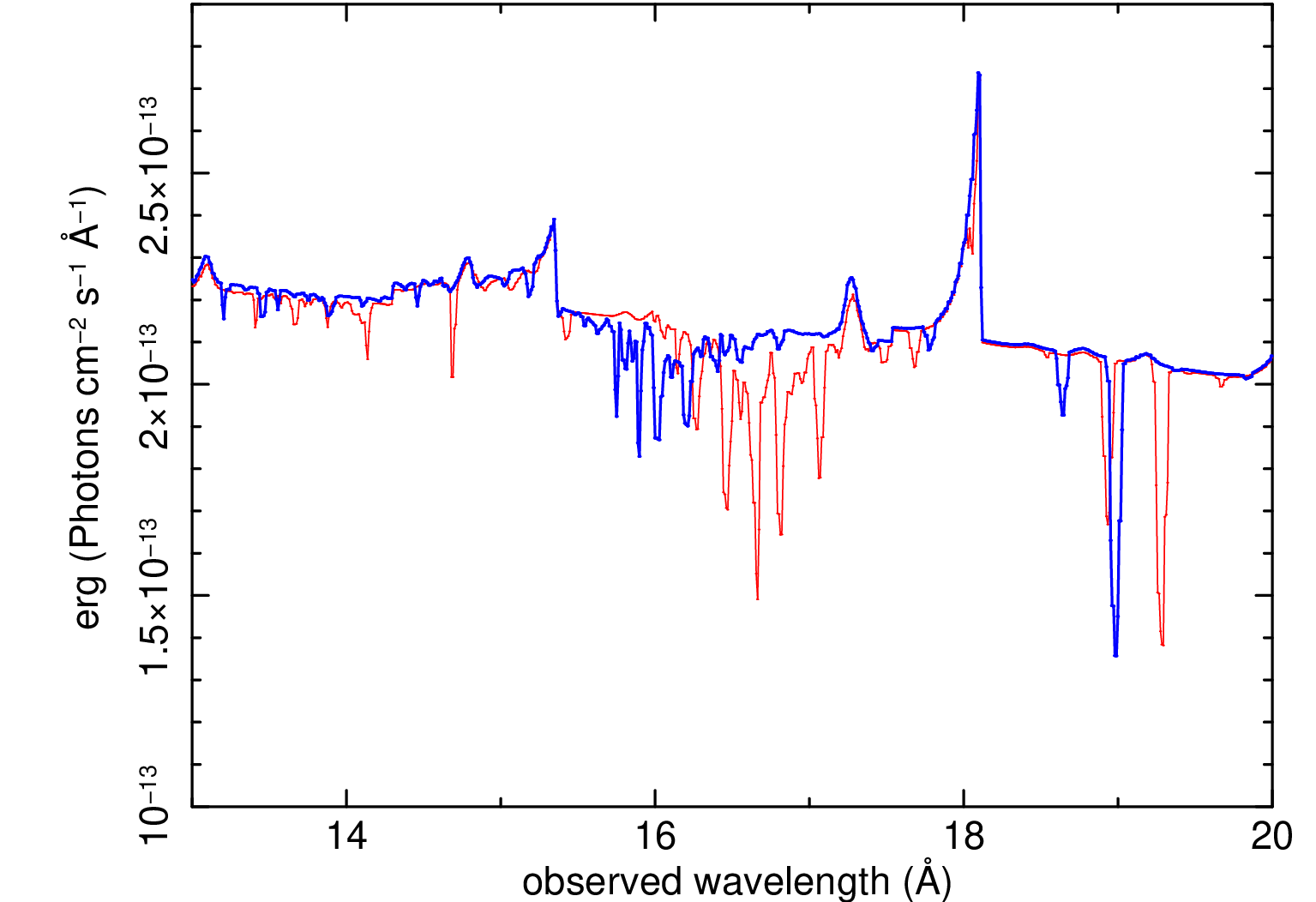}                                             
\centering 
\includegraphics[width=4.3cm, angle=270]{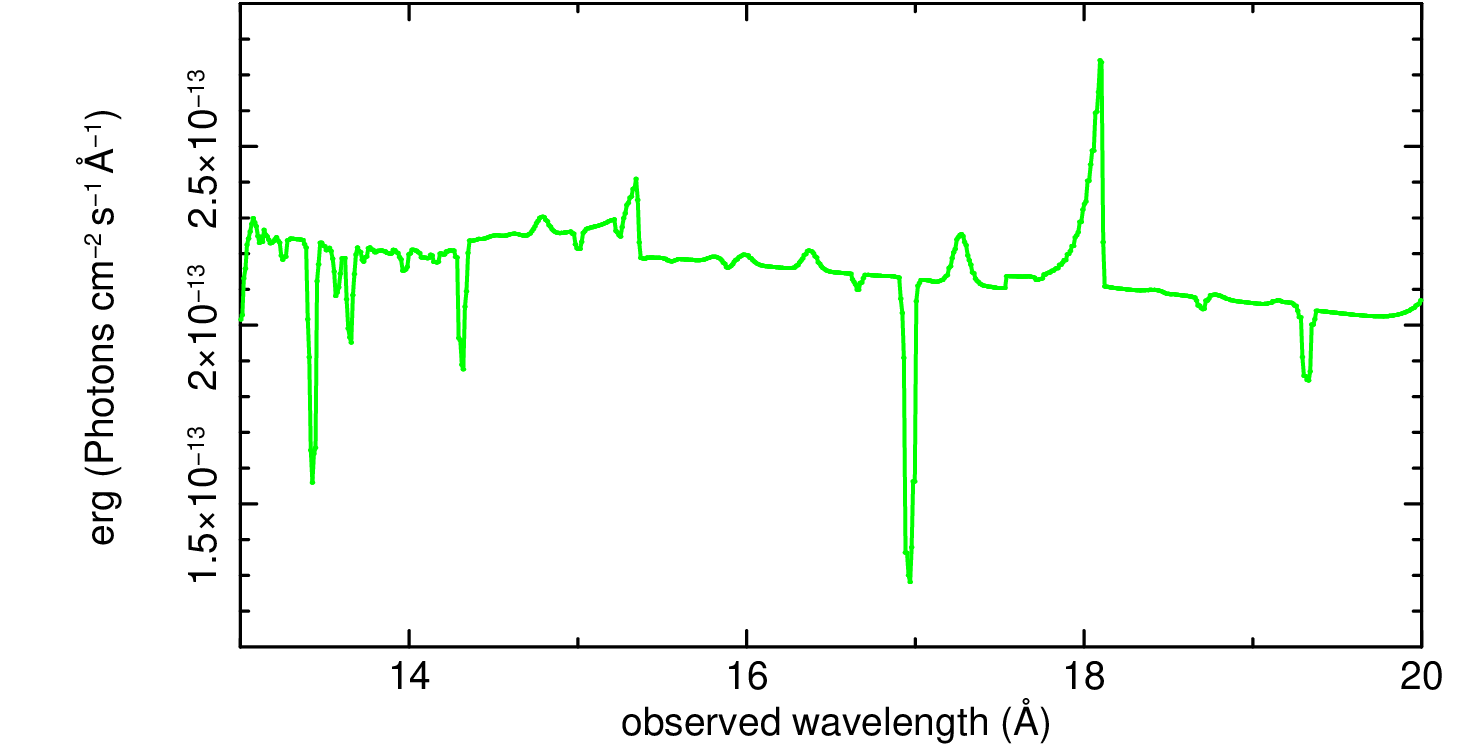}                                             
\caption                                                                
{(top) photoionized absorption spectra of the two low ionization components listed in Table 2 as abs1 and abs2, with outflow velocities of v$\sim$0.061c (red) and v$\sim$0.077c (blue).
(lower) a similar plot for the more highly ionized absorber outflowing at v$\sim$0.188c (green)}
\end{figure}

\section{Acknowledgements}
\xmm\ is a space science mission developed and operated by the European Space Agency. {We acknowledge in particular the excellent work of ESA staff at the European Space Astronomy Center in Madrid in 
successfully planning and conducting the relevant \xmm\ observations.} The UK Science and Technology Facilities Council funded the posts of AL and MC.

\end{document}